\documentstyle[aps,12pt,preprint,epsf]{revtex}

\begin{document}

\draft

\preprint{DPNU-96-20, gr-qc/9603xxx}

\title{Surface gravity in dynamical spherically symmetric spacetimes}

\author{Gyula Fodor\thanks{Present address:
Research Institute for Particle and Nuclear Physics,
H-1525 Budapest 114, P.O.Box 49, Hungary; e-mail:
gfodor@rmk530.rmki.kfki.hu},
Kouji Nakamura, Yoshimi Oshiro and Akira Tomimatsu}

\address{Department of Physics, Nagoya University, Chikusa-ku, Nagoya 464-01,
Japan}

\date{\today}

\maketitle

\begin{abstract}

A definition of surface gravity at the apparent horizon of dynamical
spherically symmetric spacetimes is proposed. It is based on a unique foliation
by ingoing null hypersurfaces. The function parametrizing the hypersurfaces can
be interpreted as the phase of a light wave uniformly emitted by some far-away
static observer.  The definition gives back the accepted value of surface
gravity in the static case by virtue of its nonlocal character. Although the
definition is motivated by the behavior of outgoing null rays, it turns out
that there is a simple connection between the generalized surface gravity, the
acceleration of any radially moving observer, and the observed frequency change
of the infalling light signal. In particular, this gives a practical and simple
method of how any geodesic observer can determine surface gravity by measuring
only the redshift of the infalling light wave. The surface gravity can be
expressed as an integral of matter field quantities along an ingoing null line,
which shows that it is a continuous function along the apparent horizon. A
formula for the area change of the apparent horizon is presented, and the
possibility of thermodynamical interpretation is discussed. Finally, concrete
expressions of surface gravity are given for a number of four-dimensional and
two-dimensional dynamical black hole solutions.

\end{abstract}


\narrowtext

\section{Introduction}

There has been a renewed interest in spherically symmetric spacetimes in the
past half decade. The unexpected complexity of the problem is well illustrated
by the large number of new `dirty' black hole solutions. An important result
proved by Visser\cite{viss} is that these stationary matter fields always
decrease the surface gravity compared to the same mass vacuum black hole.
Dynamical spherically symmetric spacetimes were initially studied mainly to
check the validity of the cosmic censorship hypothesis. More recently,
supported by powerful numerical methods, considerable efforts have been
focused on near-critical collapsing solutions at the black hole formation
threshold.

The fundamental question is what physical quantities are describing these
spherically symmetric collapses. Undoubtedly, local quantities such as a now
well defined gravitational mass function and densities belonging to the matter
fields are essential. Because of their importance in the stationary case, we
expect that generalizations of thermodynamical quantities will also play a
major role. For stationary black holes, the surface gravity is proportional to
the temperature of the Hawking radiation. On the other hand, considering the
collapse of a spherical shell, Hiscock\cite{hisc} proposed to identify
one-quarter of the area of the apparent horizon as the gravitational entropy.
Furthermore,  Hajicek\cite{haji} suggested that the Hawking effect is
associated with the apparent horizon rather than the event horizon, since the
apparent horizon in spherically symmetric spacetimes acts as the boundary of
negative energy states. Hence we expect that some naturally generalized surface
gravity for apparent horizons will have a crucial role as a physical quantity
in dynamical spacetimes.

It is possible to formulate a local dynamical analogue of black hole
thermodynamics even for not spherically symmetric apparent horizons. Using a
null tetrad formalism, Collins\cite{coll} has derived a formula for the area
change of the apparent horizon, which can be interpreted as a generalized first
law of thermodynamics. However, the temperature term in this equation is tetrad
dependent, and no unique tetrad choice is made in the nonstationary case. This
ambiguity reflects the difficulty of selecting an appropriate distance function
along the apparent horizon. Hayward\cite{hayw} uses the natural distance
defined by the spacetime metric, although this choice is divergent in the
stationary limit when the horizon becomes null. Hayward also presents the
analogues of the zeroth and second laws of thermodynamics, and defines a
dynamical counterpart of surface gravity, called trapping gravity.
Unfortunately, when specializing to static spherically symmetric spacetimes
again, the trapping gravity does not agree with the accepted value of surface
gravity even for charged Reissner-Nordstr\"om black holes (see Appendix).

There are two basic ways to introduce surface gravity in stationary spacetimes.
The first, physically more direct method is in terms of the acceleration of
stationary observers near the black hole horizon. This form of the definition
proves to be very difficult to generalize. In dynamical spherically symmetric
spacetimes, the observers moving on constant radius orbits are the most natural
analogues of the static observers. However, their acceleration have a
qualitatively different behavior, being proportional to the matter density
instead of any possible generalization of surface gravity.

The second, mathematically more straightforward approach is to define surface
gravity as the inaffinity of the Killing vector field along the black hole
horizon. In the general dynamical case the apparent horizon ceases to be null,
and there are no geodesics remaining tangent to it. However, in spherical
symmetry, the outgoing radial light rays are necessarily geodesics, and they
are locally constant radius orbits when they cross the apparent horizon. Hence
it is natural to attempt to generalize surface gravity as the inaffinity of
these outgoing null orbits\cite{cart}. The concept of inaffinity is defined
only with respect to a preferred parametrization of the curves. The main
difficulty is how to choose this particular parametrization, considering that
one has to get back the Killing time in the stationary limit. The normalization
of the Killing vector is defined at spacelike infinity, which shows that our
definition cannot be local either.

The most important idea in this paper is to parametrize the outgoing null
geodesics using a natural spherically symmetric foliation of ingoing null
hypersurfaces. We assume that the labelling of these hypersurfaces is defined
by the proper time of a static observer at infinity. This foliation can be most
easily observed by any dynamical observer, simply by observing a radio wave
emitted uniformly by a far away reference clock. We can interpret the function
generating this foliation as a global advanced-time cooridinate. We will see
that for static observers in a static spherically symmetric spacetime, this
advanced-time parameter agrees with the Killing time, (apart from a radius
dependent additive constant,) which ensures that our generalized surface
gravity indeed reduces to the accepted value in the non-dynamical case.

Our natural ingoing null foliation will allow us not only to give a clear
physical interpretation of surface gravity, but also to prescribe the most
practical way to measure it in any static or dynamical spherically symmetric
spacetime. Any observer moving along a timelike orbit can precisely measure the
apparent change of frequency of a standard radio or light signal falling in
from the far away asymptotically flat region. We can find an explicit relation
between this frequency change and the acceleration of the observer. In
particular, for any geodesic observer crossing the horizon, the proper time
derivative of the redshift of the infalling wave is exactly equal to the
surface gravity. This is particularly interesting, since it means that surface
gravity can be determined by performing simple frequency measurements only.

In principle, there are infinitely many inequivalent ways to extend the
definition of surface gravity to nonstationary spacetimes. One has to apply
physical considerations to make the most appropriate choice. Actually, most of
these potential extensions does not seem to possess any invariant physical
meaning at all. In contrast, as we will see, our definition of generalized
surface gravity is supported not only by the analysis of outgoing light rays,
but also by simple measurements performed by both geodesic and accelerating
observers.

Another physical approach, which may lead to a different (but still not local)
definition of surface gravity, is by using a fully dynamical generalization of
the Hartle-Hawking formula\cite{haha}. Assuming that the apparent horizon area
corresponds to the entropy of a dynamical black hole\cite{hisc}, this formula
may be interpreted as a generalized first law of black hole thermodynamics. It
has been pointed out by Collins\cite{coll}, that the temperature term appearing
in this equation can correspond to some possible generalization of surface
gravity only in the near-stationary limit. Furthermore, this temperature term
can change in a noncontinuous way along the horizon whenever there is a jump in
the matter field density. This happens for example at the surface of a
collapsing star. In contrast, as we will see, our generalized surface gravity
is always continuous for regular matter fields.

Our dynamical surface gravity is defined in section \ref{sec:gensf}, using the
inaffinity of outgoing null rays at the apparent horizon.  We note that the
same definition can be applied at any point of the spacetime, including the
event horizon, but the physical meaning would be much less clear there. In
section \ref{sec:measg}, a method is described for how any observer, which
crosses the horizon in an arbitrary way, can measure surface gravity by
observing light signals falling in from infinity. A family of observers whose
accelerations are equal to the surface gravity multiplied by a generalized
redshift factor is also presented. In section \ref{sec:genfl}, the surface
gravity is expressed as an integral of regular matter field quantities along an
ingoing null curve coming from past null infinity. In section \ref{sec:arla},
an equation for the area change of the apparent horizon is presented. The
possibility of interpreting it as a dynamical first law of black hole
thermodynamics is discussed.  In section \ref{sec:examp}, the value of the
surface gravity is calculated for several exact solutions. These include the
charged Vaidya metric, self-similar scalar field solutions, $1+1$ dimensional
dilaton gravity and homogenous dust ball collapse. In the Appendix, while
examining the properties of null congruences, Hayward's definition of trapping
gravity is reviewed in the spherically symmetric case, and the relation to our
formulation is discussed. We use units in which the gravitational constant and
the speed of light satisfy $G=c=1$.

\section{Generalized surface gravity}\label{sec:gensf}

The surface gravity $\kappa$ of stationary spacetimes is defined by the
behavior of the timelike Killing vector $\xi^\alpha$ at the event horizon. The
definition has a non-local character. If $\xi^\alpha$ is a Killing vector
field, then $b\xi^\alpha$ is also Killing for any constant $b$. This changes
the value of the surface gravity from $\kappa$ to $b\kappa$. Therefore one must
fix the normalization of $\xi^\alpha$. The obvious way to do it is to require
$\xi^\alpha\xi_\alpha=-1$ at spacelike infinity. To calculate surface gravity,
either one has to know the Killing vector field globally, or has to perform
integration between the horizon and spacelike infinity to determine the
`anomalous redshift factor'\cite{viss}.

There are several equivalent expressions which can be used to define surface
gravity in stationary spacetimes. The most appropriate for generalizing into
dynamical spacetimes is
\begin{equation}
\xi^\beta\xi^\alpha_{\ ;\beta}=\kappa\xi^\alpha\ . \label{stasfg}
\end{equation}
This defining equation, unlike the others, only uses the value of the Killing
vector  $\xi^\alpha$ strictly on the horizon. It describes the failure of
$\xi^\alpha$ being affine null geodesic.  Since the wave vector of a light
signal is affine null geodesic, $\kappa$ has the physical meaning of
determining the frequency decrease,  or in other words the {\em redshift}, of
an outgoing light signal moving along the horizon, measuring it  with respect
to the Killing-time. Hence $\kappa$ describes the `energy loss' of a photon
trying to climb out of the black hole, but only able to move exactly along the
constant radius horizon.  No such frequency change occurs for a light signal
moving exactly along the horizon of an extreme Reissner-Nordstr\"om black hole,
although the redshift of a photon escaping from very close to the horizon to
infinity can be still arbitrarily large.

Given any spherically symmetric spacetime which is asymptotically flat at past
null infinity, let us consider a foliation by ingoing null hypersurfaces
parametrized by a function $v$. We make the parametrization of these
spherically symmetric hypersurfaces unique (up to an additive constant) by
requiring that $\xi^\alpha v_{;\alpha}=1$ at past null infinity, where
$\xi^\alpha$ is the asymptotic Killing vector. This requirement means that $v$
is fixed by the proper time of some stationary observer at infinity. We can
consider the function $v$ as a global advanced-time coordinate. The
parametrization of the null surfaces can be more conveniently fixed using the
natural radial function $\rho$ instead of $\xi^\alpha$. At infinity
$\xi^\alpha\xi_\alpha=-1$, $\rho^{;\alpha}\rho_{;\alpha}=1$ and
$\xi^\alpha\rho_{;\alpha}=0$. Hence in place of $\xi^\alpha v_{;\alpha}=1$ we
can equivalently require $\rho^{;\alpha}v_{;\alpha}=1$ at past null infinity.

It is easy to see that in a static spherically symmetric spacetime $\xi^\alpha
v_{;\alpha}$ is constant along the ingoing constant $v$ lines:
\begin{equation}
v^{;\beta}\left(\xi^\alpha v_{;\alpha}\right)_{;\beta}=
v^{;\alpha}v^{;\beta}\xi_{(\alpha;\beta)}
+\frac12\xi^\alpha\negthinspace\left(v^{;\beta}v_{;\beta}\right)_{;\alpha}=0\ .
\end{equation}
Hence $\xi^\alpha v_{;\alpha}=1$ everywhere. This means that for static
observers the advanced time $v$ agrees with the Killing time, apart from an
observer dependent additive constant determining the time `zero'.

\begin{figure}[h]
  \begin{center}
    \leavevmode
    \epsfysize=20pc \epsfbox[132 345 492 610]{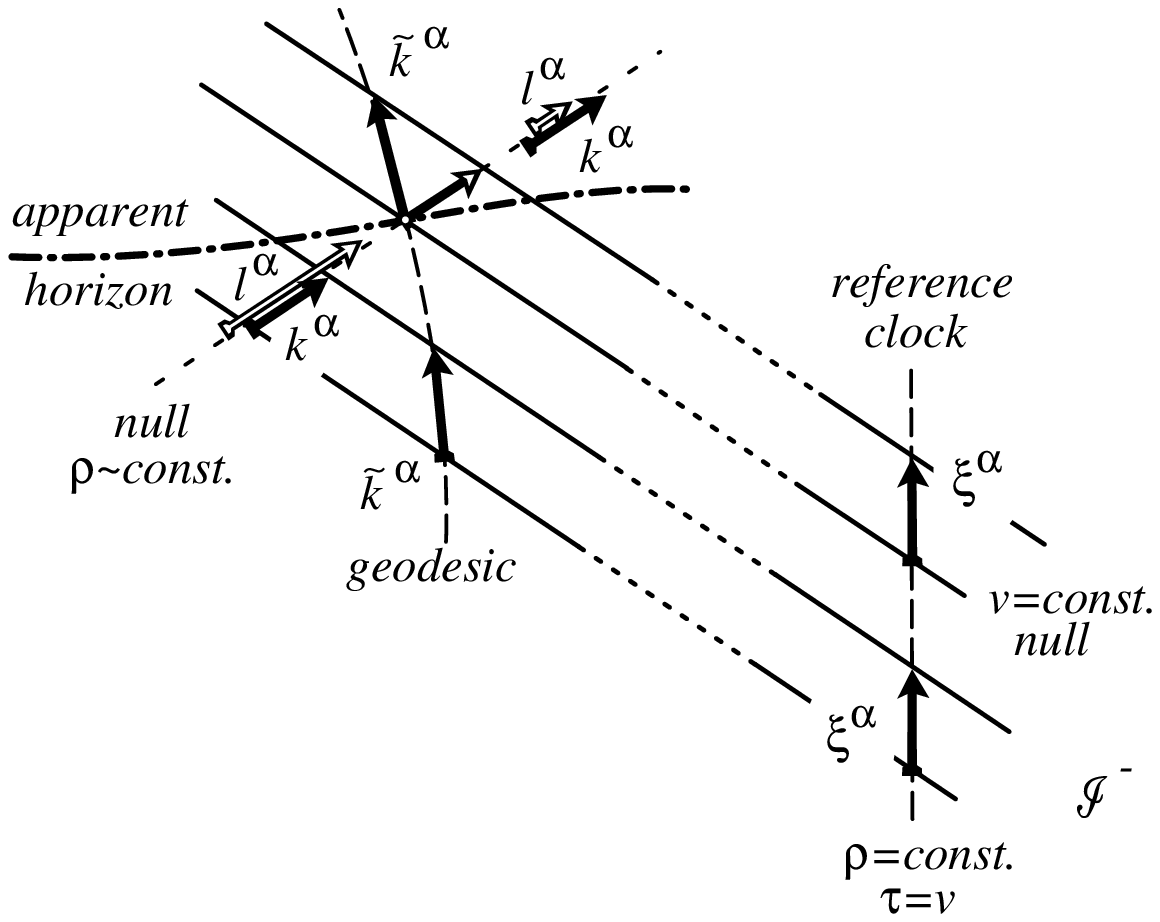}
  \end{center}
\caption{
An illustration of our dynamical surface gravity definition. The function $v$
parametrizing the null foliation  is fixed by the proper time $\tau$ of a far
away reference clock. The velocity vector of the clock agrees with the
asymptotic Killing vector $\xi^\alpha$, and $\xi^\alpha v_{;\alpha}=1$. The
surface gravity describes the inaffinity of  both $k^\alpha$ and $\tilde
k^\alpha$ (black arrows), as long as they are tangent to a geodesic and
$k^\alpha v_{;\alpha}=\tilde k^\alpha v_{;\alpha}=1$. The white arrow
represents the affine geodesic wave vector $l^\alpha$ of some outgoing light
signal crossing the horizon. The `frequency' $l^\alpha v_{;\alpha}$ is a
decreasing function of the advanced-time $v$.
}\label{figdef}
\end{figure}

In a dynamical spacetime the apparent horizon is not null anymore, and there
are no geodesics which remain tangent to it. However, outgoing radial null
curves are always geodesic because of spherical symmetry.  Furthermore, since
the expansion of outgoing null rays vanishes at the apparent horizon, the
outgoing null curves are locally constant radius orbits when they cross the
horizon. Hence instead of the Killing vector, which is very problematic to
generalize to dynamical spacetimes, we will use the inaffinity of an outgoing
null vector field $k^\alpha$ to define surface gravity. The null condition and
the spherical symmetry only fixes the direction of $k^\alpha$. The most
difficult problem is how to fix the normalization of this vector field. Since
we want our definition to give back the usual value for the surface gravity
when specializing to static spacetimes, $k^\alpha$ should agree with the
Killing vector $\xi^\alpha$ on a static horizon.  We can assure this by
requiring $k^\alpha v_{;\alpha}=1$ at every point of the spacetime. This
determines $k^\alpha$ uniquely in a non-local way. Because $k^\alpha$ is
geodesic everywhere, the relation $k^\beta k^\alpha_{\ ;\beta}=\kappa k^\alpha$
can be used to define $\kappa$ at every point of the spacetime which can be
reached by an ingoing radial lightray coming from past null infinity. However,
on physical grounds, we are interested in the value of the surface gravity only
at the apparent horizon. Since $k^\alpha\rho_{;\alpha}=0$ only on the apparent
horizon, the physical significance of $\kappa$ is much less clear elsewhere
(see Fig.\ \ref{figdef}).

Multiplying the formula  $k^\beta k^\alpha_{\ ;\beta}=\kappa k^\alpha$ by
$v_{;\alpha}$\,,
\begin{equation}
\kappa=v_{;\alpha}k^\beta k^\alpha_{\
;\beta}=k^\beta\left(v_{;\alpha}k^\alpha\right)_{;\beta}-k^\alpha k^\beta
v_{;\alpha\beta}=
-k^\alpha k^\beta v_{;\alpha\beta}\ .
\end{equation}
Since $v^{;\beta}v_{;\alpha\beta}=v^{;\beta}v_{;\beta\alpha}=0$, for any scalar
function $a$ the vector $\tilde k^\alpha=k^\alpha+a v^{;\alpha}$ will also
satisfy $\kappa=-\tilde k^\alpha\tilde k^\beta v_{;\alpha\beta}$. This shows
that the fundamental structure is not the vector field $k^\alpha$, but the
function  $v$ determining the null foliation. We only have to assume that
$k^\alpha$ points in a  radial direction and $k^\alpha v_{;\alpha}=1$. Since no
derivative of $k^\alpha$ appears, it is enough to choose any such vector at
only one point, no need to construct a vector field. The vector $k^\alpha$ can
be not only null but also timelike or spacelike.

{\bf Definition:} Given a foliation by ingoing null hypersurfaces, parametrized
by a function $v$ which satisfies $v_{;\alpha}\rho^{;\alpha}=1$ (or $\xi^\alpha
v_{;\alpha}=1$) at past null infinity, the {\em surface gravity} at some point
of the spacetime is defined as
\begin{equation}
\kappa=-k^\alpha k^\beta v_{;\alpha\beta}\ , \label{kappa}
\end{equation}
where $k^\alpha$ is a vector pointing in a radial direction and satisfying
$k^\alpha v_{;\alpha}=1$.

Given any radially directed geodesic, we can parametrize it by the advanced
time $v$. Then the tangent vector $\tilde k^\alpha$ satisfies $\tilde k^\alpha
v_{;\alpha}=1$ and the geodesic equation $\tilde k^\beta\tilde k^\alpha_{\
;\beta}=\tilde\kappa\tilde k^\alpha$. Multiplying by $v_{;\alpha}$, we get
$\tilde\kappa=\kappa$ (see Fig.\ \ref{figdef}).

{\bf Consequence:} For any radial geodesic with tangent vector $k^\alpha$
satisfying $k^\alpha v_{;\alpha}=1$, the surface gravity $\kappa$ describes the
{\em inaffinity} of the geodesic as
\begin{equation}
k^\beta k^\alpha_{\ ;\beta}=\kappa k^\alpha\ .
\end{equation}

The physically most relevant case is when $k^\alpha$ is the unique outgoing
null vector crossing the apparent horizon and satisfying $k^\alpha
v_{;\alpha}=1$. At the horizon of static black holes this null vector agrees
with the Killing vector, and our definition gives the standard value of surface
gravity. The physical meaning of the dynamical $\kappa$ is the same as in the
stationary case. An outgoing light signal moves along a locally constant radius
orbit when it crosses the apparent horizon. Since the parametrization $v$ is
not affine, $\kappa$ determines the frequency decrease, that is the {\em
redshift} of the light signal at the horizon (see Fig.\ \ref{figdef}).
Physically, the photon loses its `energy' because of the attractivity of the
black hole.

What happens if we try to calculate the surface gravity using a different
parametrization of the null hypersurfaces, a function $\tilde v$ which is not
asymptotically well behaving at past null infinity? Then we get
$\tilde\kappa=-\tilde k^\alpha\tilde k^\beta\tilde v_{;\alpha\beta}$ for some
$\tilde k^\alpha$ satisfying  $\tilde k^\alpha\tilde v_{;\alpha}=1$. The
physical parametrization always can be obtained by a relabelling of the null
surfaces, $v\equiv v(\tilde v)$. Since $\tilde k^\alpha
v_{;\alpha}=\frac{dv}{d\tilde v}\equiv v'$, we have to rescale the vector
$\tilde k^\alpha$ and use $k^\alpha=\tilde k^\alpha/v'$ to ensure that
$k^\alpha v_{;\alpha}=1$. Expressing the covariant derivative in terms of the
Christoffel symbols and partial derivatives in some coordinate system,
\begin{equation}
\kappa=-k^\alpha k^\beta v_{;\alpha\beta}=-\frac1{(v')^2}\tilde k^\alpha\tilde
k^\beta\left[\left(v'\tilde v_{,\alpha}\right)_{,\beta}
-\Gamma^\gamma_{\ \alpha\beta}v'\tilde v_{,\gamma}\right]\ .
\end{equation}
Hence the physical surface gravity $\kappa$ is related to the unphysical
$\tilde\kappa$ as
\begin{equation}
\kappa=\frac1{v'}\tilde\kappa-\frac1{(v')^2}v''
=\frac1{v'}\tilde\kappa+\left(\frac1{v'}\right)' \ , \label{trafo}
\end{equation}
where the primes denote derivatives with respect to $\tilde v$.

Given the function $v$, one can choose it as one of the coordinates in a null
coordinate system $x^\alpha=(v,r,\theta,\phi)$. The metric takes the form
\begin{equation}
ds^2=-Fdv^2+2Gdvdr+\rho^2d\Omega^2\ ,\label{nullco}
\end{equation}
where $F$, $G$ and $\rho$ are functions of $v$ and $r$, and $G>0$. The only
remaining freedom is in choosing the $r$ coordinate. Using the Christoffel
symbols in this coordinate system, from (\ref{kappa}) we have $\kappa=k^\alpha
k^\beta\Gamma^1_{\ \alpha\beta}$\,, and since $\Gamma^1_{\ 22}=\Gamma^1_{\
12}=0$,
\begin{equation}
\kappa=\Gamma^1_{\ 11}=\frac{G_{,v}}G+\frac{F_{,r}}{2G}\ , \label{kavr}
\end{equation}
independently of the radius function $\rho$. If we choose $r$ as an outgoing
null coordinate, then we obtain a double-null coordinate system with $F=0$ and
$\kappa=G_{,v}/G$. Another convenient choice is $r=\rho$, which we will use in
most of the paper.

\section{Physical implications}\label{sec:measg}

The familiar method of determining the surface gravity of a stationary black
hole is by measuring the acceleration of observers moving along the Killing
orbits near the horizon. Using the coordinate system (\ref{nullco}) where
$r=\rho$, the most natural generalization of the Killing vector is
$\xi^\alpha=\left(1,0,0,0\right)$, because it satisfies
$\xi^\alpha\rho_{;\alpha}=0$, and reduces to the Killing vector in the static
case. Since $\xi^\alpha\xi_\alpha=-F$, the velocity of the observers moving
along these constant radius orbits is  $u^\alpha=\left(1/\sqrt F,0,0,0\right)$.
Their acceleration is $a^\alpha=u^\beta u^\alpha_{;\beta}$, and
\begin{equation}
\xi^\alpha\xi_\alpha\,a^\beta
a_\beta=\left(\frac{G_{,v}}G+\frac{F_{,r}}{2G}+\frac{F_{,v}}{2F}\right)^2\ .
\end{equation}
In stationary spacetimes $F_{,v}=0$, and comparing with (\ref{kavr}) we can see
that this expression gives $\kappa^2$. In a dynamical spherically symmetric
spacetime the derivative of the radius function $\rho=r$ vanishes in outgoing
null directions at the points of the apparent horizon. Since $(2G,F,0,0)$ is
such a null vector field, this means that $F=0$ there. However, from the
Einstein's equations we have  $F_{,v}=-8\pi\rho G T_{vv}$, and hence $F_{,v}/F$
and $\xi^\alpha\xi_\alpha\,a^\beta a_\beta$ in general diverges at the apparent
horizon. The only combination which is always finite is
$\left(\xi^\alpha\xi_\alpha\right)^3a^\beta a_\beta$. It is proportional to
$T_{\alpha\beta}\xi^\alpha\xi^\beta$ instead of $\kappa$, and zero in the
static case. From these arguments we can see, that in dynamical spacetimes the
acceleration of constant radius observers cannot be used to define any
generalization of apparent horizon surface gravity. To illustrate the problem
more concretely, let us consider the Vaidya spacetime, for which $G=1$,
$F=1-2m(v)/r$ and $\rho=r$. Then $\kappa=m/r^2$, and
\begin{equation}
\xi^\alpha\xi_\alpha\,a^\beta a_\beta=\left(\frac
m{r^2}-\frac{2m'}{r-2m}\right)^2\ .
\end{equation}
This diverges at the apparent horizon $r=2m$, whenever $m'=\frac{dm}{dv}\not=0$
in the dynamical case.

There is a very intimate connection between the acceleration of radially moving
observers, the advanced time $v$ and the generalized surface gravity $\kappa$.
Consider an arbitrary congruence of curves in the constant angle radial plane,
generated by some vector field $u^\alpha$. We do not assume that the congruence
is geodesic, and it can be not only timelike but also spacelike or null. If we
define the function $f$ by $f=u^\alpha v_{;\alpha}$, the vector field
$k^\alpha=u^\alpha/f$ satisfies the normalization condition $k^\alpha
v_{;\alpha}=1$ needed in the definition of $\kappa$. We define the vector field
$a^\alpha=u^\beta u^\alpha_{;\beta}$, which is just the acceleration when
$u^\alpha$ is a normalized velocity vector. Then $a^\alpha=f^2k^\beta
k^\alpha_{;\beta}+fk^\alpha k^\beta f_{;\beta}$. Multiplying this by
$v_{;\alpha}$, using that the derivative of $k^\alpha v_{;\alpha}$ vanishes,
and substituting the defining relation (\ref{kappa}) of $\kappa$, we get
\begin{equation}
a^\alpha v_{;\alpha}=f^2\kappa+u^\alpha f_{;\alpha} \ . \label{acv}
\end{equation}

\begin{figure}[h]
  \begin{center}
    \leavevmode
    \epsfysize=15pc \epsfbox[148 395 442 604]{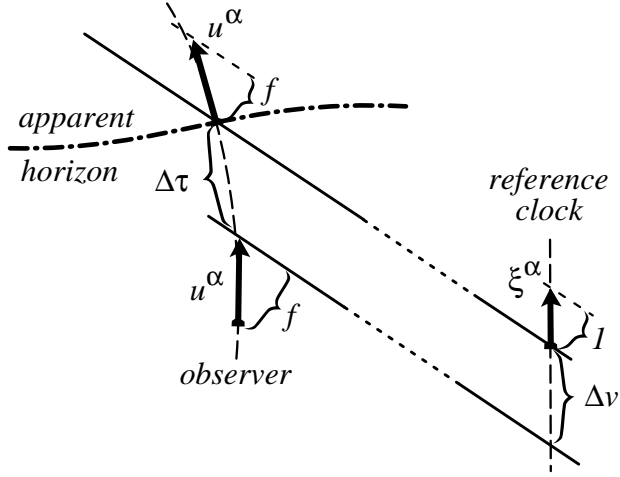}
  \end{center}
\caption{
Given an observer with velocity $u^\alpha$, define the function $f=u^\alpha
v_{;\alpha}$. If there is an infalling light wave emitted by a far-away static
clock with frequency $\omega_\infty$, the observer measures
$\omega=f\omega_\infty$ frequency. For a geodesic observer, the surface gravity
$\kappa$ is equal to the derivative of the redshift $z$ with respect to the
proper time $\tau$, i.\ e.\ $\kappa=u^\alpha z_{;\alpha}$ where $z=1/f-1$.
}\label{meas}
\end{figure}

If the congruence is geodesic, then $a^\alpha=0$ and
$\kappa=u^\alpha(1/f)_{;\alpha}$. This is not very surprising, since we have
seen in the previous section that $\kappa$ describes the inaffinity of any
geodesic parametrized by $v$. The important thing is that $f$ has a simple
physical interpretation and can be very easily measured. Since $f=u^\alpha
v_{;\alpha}$, the value of $f$ gives  the ratio of the global advanced time
change $\Delta v$ and the observer's proper time change $\Delta\tau$ along the
orbit (see Fig\ \ref{meas}):
\begin{equation}
f=\frac{d v}{d\tau}\ .
\end{equation}
Considering a light signal emitted by a static observer at infinity with
frequency $\omega_\infty$, the observed frequency is $\omega=f\omega_\infty$.
Introducing the redshift factor \begin{equation}
z=\frac{\omega_\infty}\omega-1=\frac1f-1\ ,
\end{equation}
the surface gravity is
\begin{equation}
\kappa=u^\alpha z_{;\alpha}=\frac{d z}{d\tau}\ .
\end{equation}
This shows that for any geodesic observer, the proper-time derivative of the
observed redshift of a standard light or radio signal is equal to the surface
gravity $\kappa$. Since such frequency changes can be very easily and most
precisely determined, this is the most practical method of measuring surface
gravity in spherically symmetric spacetimes, even in the static case.

The proper time is measured by a clock carried by the observer. Actually, since
we have not used that the norm of $u^\alpha$ is $-1$, $\tau$ does not even have
to be proper time, it is enough if  proportional to it. But to get the physical
$\kappa$,  the frequency $\omega_\infty$ of the light signal must be determined
by the proper time of a static clock at infinity. To measure the surface
gravity of the apparent horizon, the observer must actually cross the horizon.
If the apparent horizon is spacelike, the observer is unable to send the result
of the measurement back to infinity. The utmost an observer far from the black
hole can know is the approximate value of $\kappa$ at the event horizon, even
if the physical meaning of $\kappa$ is not clear there. However, we expect that
the generalized surface gravity will play the most important role in the case
of dynamical evaporating black hole models, when the apparent horizon is
timelike and located outside of the event horizon.

Equation (\ref{acv}) provides the most practical way of  measuring surface
gravity for non geodesic observers as well. If the norm of $u^\alpha$ is
constant along the orbits, then $u^\alpha a_\alpha=0$, and (\ref{acv})
determines the only nonvanishing component of $a^\alpha$. In our null
coordinate system let
\begin{equation}
u^\alpha_\mp=\left(f,\frac{f^2F\mp1}{2fG},0,0\right)\ , \label{vect}
\end{equation}
assuming $f\geq0$. Since $g_{\alpha\beta}u^\alpha_\mp u^\beta_\mp=\mp1$ and
$g_{\alpha\beta}u^\alpha_- u^\beta_+=0$, a general observer moving in a radial
direction can be described by the velocity $u^\alpha_-$, while $u^\alpha_+$ is
the outward pointing normal vector to the orbits. Defining
$a^\alpha_\mp=u^\beta_\mp u^\alpha_{\mp;\beta}$, we have
\begin{equation}
a^\alpha_\mp=\frac1{z+1}\left(\kappa-u^\beta_\mp z_{;\beta}\right)u^\alpha_\pm\
. \label{acc}
\end{equation}
This follows from the fact that  $a^\alpha_\mp$ has to be parallel to
$u^\alpha_\pm$, and that by contracting with $v_{;\alpha}$ we get back
(\ref{acv}). The norm of $u^\alpha_+$ is $1$, the acceleration $|a^\alpha_-|$
can be directly measured, while $f$ can be determined by observing the
frequency change of light signals falling in from infinity.

All observers who measure constant redshift $z$ have acceleration proportional
to the generalized surface gravity $\kappa$. Hence to find the natural
generalization of the static observers, one has to look for those solutions of
the equation $u^\alpha_\mp f_{;\alpha}=0$ which reduce to the Killing orbits in
the static limit. Unfortunately this equation is too difficult to solve in
general. The choice $f=1/\sqrt F$ in (\ref{vect}) gives the constant radius
observers studied in the beginning of the section. However, $F$ is constant
along the constant radius orbits only in the static case. Since $F=0$ on the
horizon, one possible candidate for the solution would be the constant $F$
orbits.  Although, in general, their acceleration is not proportional to
$\kappa$, it is instructive to study these orbits in case of the Vaidya
spacetime. Since $F=1-2m(v)/r$, from $u^\alpha_\mp F_{;\alpha}=0$ we get
\begin{equation}
\frac1{f^2_\mp}=\pm F\mp2r\frac{m'}m\ ,
\end{equation}
where $f_-$ exists only for timelike, and $f_+$ only for spacelike constant $F$
orbits. Then
\begin{equation}
u^\beta_\mp\left(\frac1{f_\mp}\right)_{;\beta}=\frac{rm''}{2rm'-mF}\ .
\end{equation}
At least for the linear Vaidya spacetime, where $m=m_1v+m_0$ for some constants
$m_1$ and $m_0$, the constant $F$ orbits have acceleration proportional to
$\kappa$. In the static limit, when $m_1\rightarrow0$, these orbits reduce to
the Killing orbits.

Although solutions of the equation $u^\alpha_\mp f_{;\alpha}=0$ always exist,
for more general spacetimes it is generally impossible find these solutions.
One has the freedom to specify the value of $f$ as initial data on some
surface, for example on the apparent horizon. A natural choice is to fix $f$ by
requiring that $u^\alpha_\mp$ has to be tangent to the horizon. In general, the
orbits determined by  $u^\alpha_\mp f_{;\alpha}=0$ will not remain tangent to
the horizon. Because the orbits never cross into the other side of the horizon,
the apparent horizon will emerge as the `envelope' of these orbits. Since the
equation of these orbits is too difficult to solve, we can consider a similar
equation instead, which gives essentially the same orbits very close to the
horizon. We solve
\begin{equation}
w^\alpha_\mp f_{;\alpha}=0\ ,
\end{equation}
where
\begin{equation}
w^\alpha_\mp=\left(f,\mp\frac1{2fG_h},0,0\right) \ , \label{ww}
\end{equation}
and $G_h$ is the value of $G$ at the point of the horizon which have the same
$v$ coordinate as the point where the vector $w^\alpha_\mp$ is considered,
$G_h(v)=G\left(v,r_h(v)\right)$. If we require that $w^\alpha_\mp$ is tangent
to the horizon, i.\ e.\ $w^\alpha_\mp F_{;\alpha}\negthinspace\mid_h=0$, then
$u^\alpha_\mp$ and $w^\alpha_\mp$ not only agree on the horizon, but even their
accelerations are the same there. Defining $\tilde a^\alpha_\mp=w^\beta_\mp
w^\alpha_{\mp;\beta}$\,, and using that $w^\alpha_\mp f_{;\alpha}=0$,
$w^\alpha_\mp F_{;\alpha}\negthinspace\mid_h=0$ and $w^\alpha\partial_\alpha
G_h\negthinspace\mid_h=w^\alpha\partial_\alpha G\negthinspace\mid_h$\,, it is
straightforward to check that on the apparent horizon
\begin{equation}
\tilde a^\alpha_\mp\negthinspace\mid_h=f\kappa w^\alpha_\pm\negthinspace\mid_h
\ .
\end{equation}
The orbits of $u^\alpha_\mp$ would satisfy this property everywhere, but since
we are interested in the value of $\kappa$ only at the apparent horizon, it is
sufficient to study the much less complicated $w^\alpha_\mp$ orbits. Since
$G_h$ depends only on $v$, the equation $w^\alpha_\mp f_{;\alpha}=0$, i.\ e.
\begin{equation}
\partial_v f\mp \frac1{2f^2G_h}\partial_r f=0
\end{equation}
can always be integrated. The general solution is
\begin{equation}
r=\mp\frac1{2f^2}\int\limits^v_0\frac1{G_h}dv+c(f)\ , \label{orb}
\end{equation}
where $c(f)$ is an arbitrary function of the parameter $f$ labelling the
orbits. The function $c(f)$ is fixed by the assumption that every orbit becomes
tangent to the horizon at some point. For every value of $v$ there is an orbit
which is tangent to the horizon at the point $(v,r_h(v))$. The value of $f$ is
set for this orbit by the condition
\begin{equation}
\frac{dr_h}{dv}=\left(\frac{w^r_\mp}{w^v_\mp}\right)_h=\mp\frac1{2f_h^2G_h}\ .
\label{drh}
\end{equation}
The function $c(f)$ can be determined from
\begin{equation}
r_h=\mp\frac1{2f^2_h}\int\limits^v_0\frac1{G_h}dv+c(f_h)\ . \label{rhh}
\end{equation}

As an example, we consider Vaidya spacetime with mass $m=m_0+m_2v^2$. The
horizon is at $r_h=2m$, and from (\ref{drh})
\begin{equation}
4m_2v=\frac1{2f_h^2}\ .
\end{equation}
Substituting into (\ref{rhh}),
\begin{equation}
c(f)=2m_0\pm\frac1{32m_2f^4}\ .
\end{equation}
Using (\ref{orb}), we get the equation of the orbits:
\begin{equation}
r-2m=\pm\frac1{2m_2}\left(\frac1{4f^2}-2m_2v\right)^2\ .
\end{equation}
The vector field $w^\alpha_\mp$ can be constructed by solving this equation for
$f$ and substituting into (\ref{ww}). For timelike horizon the orbits are
always outside, while for spacelike horizon always inside the horizon. Each
orbit becomes parallel to the horizon at a point where $1/v=8m_2f^2$, depending
on the value of $f$ labelling the orbit. In the static limit when
$m_2\rightarrow0$, the vector field $w^\alpha_\mp$ becomes tangent to the
constant radius orbits.

We have seen that there always exists a family of observers instantaneously
tangent to the horizon, such that their acceleration at the moment of touching
the horizon is $f_h\kappa$. In dynamical spherically symmetric spacetimes these
orbits seem to be the most appropriate generalization of the Killing orbits
used to define surface gravity in the stationary case. $f_h$ can be interpreted
as a generalized redshift factor. It is finite for dynamical spacetimes, but
always diverges in the static limit.

\section{Integral formula}\label{sec:genfl}

In the coordinate system $ds^2=-Fdv^2+2Gdvdr+r^2(d\theta^2+\sin^2\theta
d\varphi^2)$, the independent components of the Einstein's equations are
\begin{equation}
8\pi T_{rr}=\frac{2G_{,r}}{rG}\ ,\label{err}
\end{equation}
\begin{equation}
8\pi T_{vr}=-\frac{2G}{r^2}M_{,r}\ ,\label{evr}
\end{equation}
\begin{equation}
8\pi T_{vv}=\frac2{r^2}\left(FM_{,r}+GM_{,v}\right)\ ,\label{evv}
\end{equation}
\begin{equation}
8\pi\left(T_{\theta\theta}-\frac{r^2}4T^{\
\alpha}_\alpha\right)=\frac{r^2}{2G}\kappa_{,r}+\frac Mr\ ,\label{ett}
\end{equation}
where
\begin{equation}
M=\frac r2\left(1-\frac F{G^2}\right)\ ,\qquad
\kappa=\frac{G_{,v}}G+\frac{F_{,r}}{2G}\ , \label{locm}
\end{equation}
and $T_{\alpha\beta}$ is the stress tensor of matter fields. Using the radius
function $\rho$, the local mass $M$ can be expressed in a coordinate system
invariant form as $M=\rho(1-\rho^{;\alpha}\rho_{;\alpha})/2$.

Defining the vectors $\xi^\alpha=(1,0,0,0)$ and $\ell^\alpha=(0,-1,0,0)$, we
have $T_{vv}=\xi^\alpha\xi^\beta T_{\alpha\beta}$,
$T_{vr}=-\xi^\alpha\ell^\beta T_{\alpha\beta}$ and
$T_{rr}=\ell^\alpha\ell^\beta T_{\alpha\beta}$. The vector $\ell^\alpha$ can be
easily constructed in any coordinate system, since it is future directed null
and $\ell^\alpha\rho_{;\alpha}=-1$. However, for $\xi^\alpha$, only its
direction is fixed locally by $\xi^\alpha\rho_{;\alpha}=0$, the normalization
$\xi^\alpha\xi_\alpha=-F$ is known only after globally constructing the
asymptotically well behaving foliation given by $v$. The best one can do
locally is to define the Kodama vector\cite{koda} $\zeta^\alpha$ by
$\zeta^\alpha\rho_{;\alpha}=0$ and
$\zeta^\alpha\zeta_\alpha=\frac{2M}\rho-1=\frac F{G^2}$. Then
$\xi^\alpha=G\zeta^\alpha$. Since the cotangent vector
$\rho_{;\alpha}=(0,1,0,0)$ can be easily constructed in any coordinate system,
it is useful to write the more covariant forms instead of (\ref{evr}) and
(\ref{evv}):
\begin{equation}
8\pi T_r^{\ r}=\frac{F_{,r}}{rG^2}-\frac{2M}{r^3}\ ,\label{edu}
\end{equation}
\begin{equation}
8\pi
T^{rr}=-\frac{F_{,v}}{rG^3}+\frac{2F}{rG^3}\left(\kappa-\frac{GM}{r^2}\right)\
{}.
\end{equation}

Using (\ref{err}), and assuming that $G$ approaches 1 at past null infinity,
$G$ can be expressed as an integral of local quantities along an ingoing radial
null line:
\begin{equation}
\ln G=-4\pi\int^\infty_r\limits rT_{rr}dr\ .\label{gint}
\end{equation}
This shows that $G=1$ in the whole outer vacuum region. If the matter fields
satisfy the weak energy condition, then $G$ is non-increasing in ingoing null
directions, and $0<G\leq1$. After calculating $G$, even $F$ can be determined
locally from the expression (\ref{locm}) of the local mass, $F=G^2(1-2M/r)$.

{}From equation (\ref{ett}), $\kappa$ can be expressed as an integral along an
ingoing radial null curve
\begin{equation}
\kappa=\int^\infty_r\limits
G\left(\frac{2M}{r^3}-\frac{16\pi}{r^2}T_{\theta\theta}
+4\pi T^{\ \alpha}_\alpha\right)dr\ . \label{intka}
\end{equation}
Since $G$ is not a local quantity, before calculating $\kappa$ one has to
evaluate the integral (\ref{gint}) to get $G$ at every point of the null line.
Using the expression for $T_{vr}$ from
\begin{equation}
T^{\ \alpha}_\alpha=\frac2GT_{vr}+\frac
F{G^2}T_{rr}+\frac2{r^2}T_{\theta\theta}\ ,\label{taa}
\end{equation}
we can get
\begin{equation}
\left(G\frac M{r^2}\right)_{,r}=2G\left(\pi T_{rr}-\pi T^{\ \alpha}_\alpha
+\frac{2\pi}{r^2}T_{\theta\theta}-\frac{M}{r^3}\right)\ ,
\end{equation}
which gives another integral formula for $\kappa$:
\begin{equation}
\kappa=G\frac M{r^2}+\int^\infty_r\limits 2\pi G\left(T_{rr}+T^{\
\alpha}_\alpha
-\frac6{r^2}T_{\theta\theta}\right)dr\ .
\end{equation}
This latter expression is especially useful if there are vacuum regions.

Using equation (\ref{edu}),
\begin{equation}
\frac{F_{,r}}{2G}=G\left(4\pi r T_r^{\ r}+\frac M{r^2}\right)\ .\label{frkg}
\end{equation}
The best we can do for the $G_{,v}/G$ term in the expression (\ref{kavr}) of
$\kappa$ is to take the derivative of  (\ref{gint}). We obtain
\begin{equation}
\kappa=G\left(4\pi r T_r^{\ r}+\frac M{r^2}\right)
-4\pi\int^\infty_r\limits r\frac{\partial T_{rr}}{\partial v}dr\ . \label{kadt}
\end{equation}
After $G$ is already known, the partial derivative can be written in a
coordinate system invariant form, as a Lie-derivative along the vector field
$\xi^\alpha=G\zeta^\alpha$:
\begin{equation}
\frac{\partial T_{rr}}{\partial v}=\ell^\alpha\ell^\beta{\cal
L}_{G\zeta}T_{\alpha\beta}\ .
\end{equation}
In the static case $\xi^\alpha$ is the Killing vector, and the integral term
vanishes.

It is instructive to introduce a basis carried by observers moving along the
constant radius orbits. Setting $n_\alpha=\frac G{\sqrt F}\rho_{;\alpha}$ and
$t^\alpha=\frac1{\sqrt F}\xi^\alpha$, we have $n^\alpha n_\alpha=1$, $u^\alpha
u_\alpha=-1$ and $u^\alpha n_\alpha=0$. This basis is valid only outside of the
horizon. The measured energy density is $\varepsilon=T_{\alpha\beta}t^\alpha
t^\beta$, the radial energy flow is  $S=T_{\alpha\beta}t^\alpha n^\beta$, and
the radial pressure is  $P=T_{\alpha\beta}n^\alpha n^\beta$. We have the
ingoing null vector $t^\alpha-n^\alpha=\frac{\sqrt F}G \ell^\alpha$. Then
\begin{equation}
T_{rr}=T_{\alpha\beta}\ell^\alpha\ell^\beta=\frac{G^2}F
T_{\alpha\beta}(t^\alpha-n^\alpha)(t^\beta-n^\beta)
=\frac{\varepsilon-2S+P}{1-\frac{2M}r}\ .
\end{equation}
This shows that $\varepsilon-2S+P$ must approach zero at the horizon.
Similarly we can get $T_{vr}=G(S-\varepsilon)$, $T_{vv}=F\varepsilon$, $T_r^{\
r}=P-S$ and $T^{rr}=P\left(1-\frac{2M}r\right)$.
Substituting into (\ref{gint})
\begin{equation}
\ln G=-4\pi\int^\infty_r\limits \frac{r(\varepsilon-2S+P)}{1-\frac{2M}r}dr\ .
\end{equation}
In the static case $S=0$, and this reduces to the formula determining the
`anomalous redshift' $\phi=-\ln G$ given by Visser\cite{viss}.  From
(\ref{kadt}), using that on the horizon $r=2M$ and $\varepsilon-2S+P=0$, we get
that the surface gravity of static black holes is\cite{viss}
\begin{equation}
\kappa=\frac G{2r}(1-8\pi r^2\varepsilon)\ .
\end{equation}

\section{Area law}\label{sec:arla}

Trying to obtain a dynamical analogue of the second law of thermodynamics, we
calculate the advanced time derivative of the apparent horizon area. The radius
and the local mass of the apparent horizon is related by $r_h=2M_h$. Since
\begin{equation}
\frac{dr_h}{dv}=2\frac{dM_h}{dv}=2\left(M_{,v}+M_{,r}\frac{dr_h}{dv}\right)\ ,
\end{equation}
we have
\begin{equation}
\frac{dr_h}{dv}=\frac{2M_{,v}}{1-2M_{,r}}\ . \label{drv}
\end{equation}
Since at the horizon $F=0$ and $T_{vr}=GT_r^{\ r}$, from (\ref{evr}) and
(\ref{evv}) we get
\begin{equation}
M_{,r}=-4\pi r_h^2T_r^{\ r}\ \ , \qquad M_{,v}=\frac{4\pi r_h^2}{G_h}T_{vv}\ .
\label{mrmv}
\end{equation}
The change of the horizon area is
\begin{equation}
\frac{dA_h}{dv}=8\pi r_h\frac{dr_h}{dv}=\frac{4\pi r_h^2\ 8\pi T_{vv}}
{G_h\left(4\pi r_hT_r^{\ r}+\frac1{2r_h}\right)} \ \ .
\end{equation}
Using (\ref{kadt}), we get
\begin{equation}
\Theta\frac{dA_h}{dv}=4\pi r_h^2T_{vv}\ ,\label{fila}
\end{equation}
where
\begin{equation}
\Theta=G_h\left(4\pi r_hT_r^{\ r}+\frac1{2r_h}\right)=
\frac1{8\pi}\left(\kappa+4\pi\int^\infty_{r_h}\limits r\frac{\partial
T_{rr}}{\partial v}dr\right). \label{temp}
\end{equation}
We can also see from (\ref{frkg}), that in the null coordinate system where
$\rho=r$, we have $\Theta=\frac{F_{,r}}{2G}$. In the quasi-stationary limit the
integral term becomes negligible, and we obtain an expression corresponding to
the Hartle-Hawking formula\cite{haha}. If we identify one-quarter of the
apparent horizon area as the gravitational entropy\cite{hisc}, then we may
interpret equation (\ref{fila}) as a generalized first law of black hole
thermodynamics.

One of the problems with the temperature term $\Theta$ is that it can be a
non-continuous function along the apparent horizon if there is a sudden change
in the matter density. Whenever there is a jump in $T_{rr}$, the integrand in
(\ref{temp}) becomes unbounded, and $\Theta$ stops being continuous too. On the
other hand, since every quantity remains regular in the integral form
(\ref{intka}) of $\kappa$, our generalized surface gravity is always continuous
when the energy densities are bounded. Another difficulty is that $\Theta$ is
not necessarily positive. Since the radius function $\rho$ is always constant
in the outgoing null direction at the horizon, $\frac{dr_h}{dv}$ and hence
$\frac{dA_h}{dv}$ is always positive for spacelike, and negative for timelike
apparent horizons. If the weak energy condition holds, then $T_{vv}\geq0$ on
the right hand side of (\ref{fila}), and $\Theta\geq0$ for spacelike while
$\Theta\leq0$ for timelike horizons. Furthermore, under the weak energy
condition, spacelike horizons are outer, while timelike horizons are inner
according to the classification of Hayward\cite{hayw} (see Appendix). Hence one
would expect that horizons, separating an asymptotically flat region from the
black hole region, are always spacelike. However, as we will see in the next
section when studying the example of pressureless dust collapse, near the
center the apparent horizon can become timelike (see Fig.\ \ref{figpor}). This
timelike region is separated form the outer spacelike region by points where
the horizon is ingoing null. In black hole evaporation models the energy
condition is violated, and if $T_{vv}<0$ then $\Theta>0$ for timelike apparent
horizons.

One would expect that the change of the black hole mass appears in the first
law of thermodynamics. Instead, the right hand side of (\ref{fila}) describes
the ingoing energy flux across the apparent horizon. Unfortunately, there is no
direct relation between this energy flux and the change of the local mass along
the horizon. Because $r_h=2M_h$ always holds on the horizon,
\begin{equation}
\frac1{8\pi}\,\frac1{2r_h}\,\frac{dA_h}{dv}=\frac{dM_h}{dv}\ ,
\end{equation}
which is independent of the surface gravity. On the other hand, the local mass
at past null infinity can be constant even if there is some non-massless matter
falling in along timelike orbits.

{}From (\ref{evv}) we can see that the horizon value of $T_{vv}$ is
proportional to the derivative of the local mass in the constant radius
outgoing null direction. Hence
\begin{equation}
\Theta\frac{dA_h}{dv}=G_h M_{,v}\ .
\end{equation}
Apart from the $G_h$ correction factor, this appears to be more similar to the
expected form of the first law, but unfortunately the derivatives in the two
sides of the equation are taken in different spacetime directions. The
derivative of the horizon area is calculated along a vector tangent to the
horizon:
\begin{equation}
\frac{dA_h}{dv}=y^\alpha A_{h;\alpha}\ \ ,\qquad
y^\alpha=\left(1,\frac{dr_h}{dv},0,0\right)\ .
\end{equation}
In general, the norm $y^\alpha y_\alpha$ is not constant. When the horizon is
spacelike, there is a unique outgoing unit-vector $z^\alpha$ tangent to the
horizon. Since $F=0$ at the horizon, using (\ref{drv}),
\begin{equation}
z^\alpha=\sqrt{\frac{1-2M_{,r}}{4G_h M_{,v}}}\  y^\alpha\ .
\end{equation}
Although this is a natural local specification of distance along the horizon,
it has the disadvantage of diverging in the static limit. Using (\ref{kadt}),
(\ref{mrmv}) and (\ref{temp}) we get
\begin{equation}
\sqrt{\Theta}\ z^\alpha A_{h;\alpha}=4\pi r_h\sqrt{\frac{r_h}{G_h}T_{vv}}\ .
\end{equation}
We can also write the right hand side into the form $2\sqrt{\pi r_h M_{,v}}$\ .
The fact that the temperature term appears under a square root follows from the
unnatural normalization of the vector $z^\alpha$. Substituting from eq.\
(\ref{conn}) in the Appendix, we get the form of the first law given by
Hayward\cite{hayw}.

\section{Examples}\label{sec:examp}

The simplest spherically symmetric dynamical spacetime for which we can
calculate the generalized surface gravity is the charged Vaidya
metric\cite{vaid}, describing a massless, charged null fluid falling into a
charged black hole. In the coordinate system (\ref{nullco}) we have $G=1$,
$F=1-2m(v)/r+e(v)^2/r^2$ and $\rho=r$, and it follows from (\ref{kavr}) that
$\kappa=m/r^2-e^2/r^3$. Since the radius of the outer and inner apparent
horizons is $r_\pm=m\pm\sqrt{m^2-e^2}$, the horizon surface gravity is
\begin{equation}
\kappa_\pm=\frac1{2r_\pm}\left(1-\frac{e^2}{r_\pm^2}\right)
=\pm\frac1{r_\pm^2}\sqrt{m^2-e^2}\ , \label{reno}
\end{equation}
in agreement with the Reissner-Nordstr\"om value. This local agreement is due
to the fact that $G=1$ and the $G_{,v}/G$ term is the only one in the
expression (\ref{kavr}) of $\kappa$ from where $v$ derivatives could appear. We
can also see that the surface gravity is always positive for the outer and
negative for the inner horizon. Taking the partial derivatives of
(\ref{reno}),
\begin{equation}
\frac{\partial\kappa_\pm}{\partial e}=\frac{\mp e^3}{r_\pm^4\sqrt{m^2-e^2}}\ ,
\ \ \
\frac{\partial\kappa_-}{\partial m}=-\frac1{r_-^2}\left(2+\frac
m{\sqrt{m^2-e^2}}\right)\ ,
\end{equation}
\begin{equation}
\frac{\partial\kappa_+}{\partial m}=\frac{4e^2-3m^2}{r_+^2\sqrt{m^2-e^2}
\left(m+2\sqrt{m^2-e^2}\right)}\ .
\end{equation}
This shows that charging this type of black holes always decreases their
outer-horizon surface gravity. If the infalling matter is not charged and
satisfies the energy conditions, then $\frac{\partial m}{\partial v}\geq0$, and
the inner-horizon surface gravity is always a decreasing function of time. The
outer-horizon surface gravity also decreases for not very strongly charged
black holes which satisfy $4e^2<3m^2$. However, if we interpret the solutions
with $\frac{\partial m}{\partial v}<0$ as black hole evaporation models, the
surface gravity of these not strongly charged evaporating black holes
necessarily increases until their mass reduces to $3m^2=4e^2$.

Our next example is the Roberts solution\cite{robe}, describing the
self-similar collapse of a massless scalar field. The solution can be most
conveniently given in the double null form $ds^2=-hdudv+\rho^2d\Omega^2$, where
$h=1$ and $\rho^2=\left[(1-p^2)v^2-2vu+u^2\right]/4$. For parameter values
$p>1$, this solution describes the formation of an unboundedly increasing mass
black hole, with a spacelike apparent horizon at $u=(1-p^2)v$. Since from eq.\
(\ref{kavr}) we have $\kappa=h_{,v}/h=0$ everywhere, the surface gravity is
zero all along the apparent horizon. This indicates that the Roberts solution
describes an extreme black hole, analogously to the $e=m$ Reissner-Nordstr\"om
metric.

However, not all self-similar black holes has vanishing surface gravity. There
is a conformally coupled scalar counterpart of the Roberts solution\cite{olch}.
The two metrics are related by a conformal transformation: $d\tilde s^2=-\tilde
hdudv+\tilde\rho^2d\Omega^2$, where $\tilde\rho^2=\tilde h \rho^2$,
\begin{equation}
\tilde h=\frac14\left(\tilde M^{\frac1{2\sqrt3}}+\tilde
M^{\frac{-1}{2\sqrt3}}\right)\ \ ,
\qquad \tilde M=\frac{u-(1+p)v}{u-(1-p)v}\ .
\end{equation}
The new apparent horizon is determined by $\tilde\rho_{,v}=0$. There, using
(\ref{kavr}),
\begin{equation}
\kappa=\frac1{\tilde h}\tilde
h_{,v}=\frac{\rho^2}{\tilde\rho^2}\left(\frac{\tilde\rho^2}{\rho^2}\right)_{,v}
=-\frac1{\rho^2}\left(\rho^2\right)_{,v}
=\frac1{2\rho^2}\left[u-(1-p^2)v\right]\ .
\end{equation}
The horizon exists for $p>1$, and it is a self-similarity line given by
$u=c(1-p^2)v/4$, where $c$ is a constant weakly depending on $p$:
$2.535<c<2.6667$. The positivity of $c$ shows that the apparent horizon is a
spacelike hypersurface. Substituting into the surface gravity formula,
\begin{equation}
\kappa=\frac{8(4-c)}{v\left[8(c-2)+(p^2-1)c^2\right]} >0\ .
\end{equation}
After the moment of black hole formation, the surface gravity gradually
decreases to zero from an infinitely big initial value, as the mass increases
unboundedly.

Our third example is the 1+1 dimensional dilaton gravity proposed by Callan,
Giddings, Harvey and Strominger\cite{cghs}. It is defined by the action
\begin{equation}
S=\frac1{2\pi}\int
dx^2\sqrt{-g}\left\{e^{-2\phi}\left[R+4(\nabla\phi)^2+4\lambda^2\right]
-\frac12\sum^N_{i=1}(\nabla f_i)^2\right\} \ ,
\end{equation}
where $g_{\alpha\beta}$ is the two-dimensional metric, $\phi$ is the dilaton
field, $f_i$ are matter fields, and $\lambda$ is a cosmological constant. It is
convenient to use a double-null coordinate system $(x^+,x^-)$, and denote the
only nonvanishing component of the metric by $g_{+-}=-\frac12 e^{2\rho}$. (Now
$\rho$ does not denote the radius function as in other parts of the paper.)
Considering $\frac1\lambda e^{-\phi}$ as a radius function, one can interpret
the solutions of this two-dimensional theory as four-dimensional spherically
symmetric spacetimes with metric
$ds^2=-e^{2\rho}dx^+dx^-+\frac1{\lambda^2}e^{-2\phi}d\Omega^2$. At the points
of the apparent horizon  $\frac1{\lambda^2}e^{-2\phi}$ is locally constant in
the outgoing null direction, that is $\partial_+\phi=0$. It follows from the
field equations that $\partial_+\partial_-(\rho-\phi)=0$, and hence by
rescaling the coordinates, one can always introduce a coordinate system where
$\rho=\phi$. The existence of this coordinate system actually follows from the
traceless nature of the two-dimensional stress tensor belonging to the matter
fields $f_i$.

The static vacuum solution of the model is given by
$e^{-2\rho}=e^{-2\phi}=\frac m\lambda-\lambda^2x^+x^-$, where $m$ is a
parameter giving the mass of the black hole. The asymptotically normed Killing
vector of this solution is $\xi^\alpha=(\lambda x^+,-\lambda x^-)$, and the
horizon is at $x^-=0$. Using the relation (\ref{stasfg}) defining the surface
gravity of stationary black holes, we get $\kappa=\lambda$, independently of
the black hole mass. We will see shortly, that our generalized surface gravity
always agrees with the cosmological constant, even for nonvacuum dynamical
solutions. The parametrization of the null foliation given by $x^+$ is not well
behaving at past null infinity, since $\xi^\alpha\nabla_\alpha x^+=\lambda x^+$
is not constant there. Introducing another parametrization $v$ defined by
$e^{\lambda v}=\lambda  x^+$, we have $\xi^\alpha\nabla_\alpha v=1$. Hence this
new parametrization is the one appearing in our definition of generalized
surface gravity. Suppose that we are given any asymptotically flat nonvacuum
dynamical solution of the field equations. There always exists a coordinate
system $(x^+,x^-)$, in which $\rho=\phi$. We assume that the solution
approaches the static vacuum solution at large distances, which means that the
parametrization defined by $e^{\lambda v}=\lambda  x^+$ has to be
asymptotically well behaving for any solution. In the $(v,x^-)$ coordinate
system $\phi$ is the same as before, but the metric components are different:
\begin{equation}
g_{v-}=\frac{dx^+}{dv}g_{+-}=e^{\lambda v}g_{+-}\ .
\end{equation}
Since on the apparent horizon $\partial_+\phi=0$, that is $\partial_v
g_{+-}=0$, from (\ref{kavr}) the generalized surface gravity is
\begin{equation}
\kappa=\frac1{g_{v-}}\partial_v g_{v-}=\lambda\ ,
\end{equation}
for any dynamical solution of the theory.

There exists a semiclassical model proposed by Russo, Susskind and
Thorlacius\cite{rust}, which reduces to the previously discussed CGHS model at
the classical level. It is defined by the one-loop effective action
\begin{eqnarray}
&&\tilde S=\frac1\pi\int dx^2\biggl[e^{-2\phi}\left(2\partial_+\partial_-\rho
-4\partial_+\phi\partial_-\phi
+\lambda^2e^{2\rho}\right)\nonumber\ \ \ \ \ \\
& &+\frac12\sum^N_{i=1}\partial_+f_i\partial_-f_i-\left(\frac N{12}-2\right)
\left(\partial_+\rho\partial_-\rho+\phi\partial_+\partial_-\rho\right)\biggr]\
{}.
\end{eqnarray}
Similarly to the previous model, the existence of a coordinate system
$(x^+,x^-)$ where $\rho=\phi$ follows from the field equations. The vacuum
solution has the form $e^{-2\rho}=e^{-2\phi}=-\lambda^2x^+x^-$, again. By the
same argument as in the previous paragraph, we can show that any asymptotically
flat dynamical black hole solution of this theory has constant surface gravity,
$\kappa=\lambda$, on the apparent horizon.

As our last example, we calculate the apparent horizon surface gravity of a
uniform density dust ball collapsing from the rest. The internal region is
equivalent to a part of the Friedmann cosmology,
\begin{equation}
ds^2=a^2\left(-d\eta^2+d\psi^2+\sin^2\psi d\Omega^2\right) \ ,
\end{equation}
where
\begin{equation}
a=c\left(\cos\frac\eta2\right)^2
\end{equation}
and $c$ is some constant. Since we match to a Schwarzschild solution at the
worldline of a dust particle at $\psi=\psi_0$, the coordinate values are
restricted into $-\pi<\eta<\pi$ and $0<\psi<\psi_0<\pi/2$. The proper time is
$\tau=c(\eta+\sin\eta)/2$. Since the radius function is $\rho=a\sin\psi$, the
local mass is given by
\begin{equation}
M=\frac\rho2\left(1-\rho^{;\alpha}\rho_{;\alpha}\right)=\frac c2\sin^3\psi \ .
\end{equation}
This shows that the mass parameter of the external solution is $m=\frac
c2\sin^3\psi_0$. If we denote the maximal radius of the ball by $r_0$, then
$r_0=c\sin\psi_0$\,, and
\begin{equation}
\frac{2m}{r_0}=\sin^2\psi_0\ ,\qquad c=\sqrt{\frac{r_0^3}{2m}}\ .
\end{equation}
Since at the apparent horizon $\rho=2M$, for $\eta>0$ the horizon is
represented by the timelike surface $\eta=\pi-2\psi$. This timelike horizon is
a future inner trapping  horizon in Hayward's classification\cite{hayw} (see
Appendix). Introducing $\tilde v=\eta+\psi$ as a null coordinate, we have
$F=G=a^2$ in (\ref{nullco}), and using (\ref{kavr}), the surface gravity
belonging to this parametrization can be calculated as
\begin{equation}
\tilde\kappa=\frac1a\left(a_{,\psi}+2a_{,\tilde v}\right)=-\cot\psi\ .
\end{equation}
Unfortunately, the parametrization $\tilde v$ is not asymptotically well
behaving when continued into the vacuum region. Hence we will have to use the
transformation formula (\ref{trafo}) to get the physical surface gravity
$\kappa$, where $v=t+r+2m\ln\left(\frac r{2m}-1\right)$ is the regular null
coordinate in the external Schwarzschild region. Looking from the vacuum
region, the matching boundary is generated by radial null geodesics with
maximal radius $r_0$. The radius and proper time of such geodesic can be
expressed using a parameter $\eta$ as
\begin{equation}
r=r_0\left(\cos\frac\eta2\right)^2\ ,\qquad
\tau=\frac12\sqrt{\frac{r_0^3}{2m}}(\eta+\sin\eta)\ .
\end{equation}
Since $r$ and $\tau$ has to agree at both sides of the boundary, the parameter
$\eta$ also agrees with the inner time coordinate $\eta$. At the boundary
$\tilde v=\eta+\psi_0$, and hence
\begin{equation}
v'\equiv\frac{dv}{d\tilde v}=\frac{dv}{d\eta}=\frac{dv}{dr}\frac{dr}{d\eta}+
\frac{dv}{dt}\frac{dt}{d\tau}\frac{d\tau}{d\eta}
=\frac{r_0\left(\cos\frac\eta2\right)^3}
{\sin\psi_0\cos\left(\psi_0-\frac\eta2\right)} \ .
\end{equation}
After taking the derivative of the reciprocal of this expression, we are ready
to substitute into (\ref{trafo}). Here $\eta$ is the coordinate value where the
constant $\tilde v$ null line crosses the boundary. Since $\tilde
v=(\pi-2\psi)+\psi$ at the horizon and $\tilde v=\eta+\psi_0$ at the matching
surface, we have $\eta=\pi-\psi-\psi_0$. The final result for the surface
gravity is
\begin{equation}
\kappa=\frac{\sin\psi_0}{4r_0\sin\psi\left(\sin\frac{\psi+\psi_0}2\right)^4}
\left[\cos(\psi-\psi_0)+2\cos(2\psi_0)-3\cos(\psi+\psi_0)\right] \ .
\end{equation}
Here, the radius at the point of the horizon where $\kappa$ is calculated is
\begin{equation}
r=c\sin^3\psi=\sqrt{\frac{r_0^3}{2m}}\sin^3\psi\ ,
\end{equation}
and $\sin^2\psi_0=2m/r_0$. At the surface of the ball, where $\psi=\psi_0$, the
the surface gravity becomes $\frac1{4m}$, which shows that our generalized
surface gravity indeed changes continuously along the apparent horizon. Close
to the central singularity, when $r$ and $\psi$ is small,
\begin{equation}
\kappa\approx-\frac2\psi\cot\frac{\psi_0}2(1+2\cos\psi_0)<0\ ,
\end{equation}
the surface gravity diverges to minus infinity.

\section{Summary and discussion}

In this paper we have proposed a generalized definition of surface gravity on
the apparent horizon of spherically symmetric dynamical black holes. Since in
stationary spacetimes the surface gravity is not a local quantity, our
definition cannot be local either. The necessary non-local structure is an
asymptotically regular foliation by ingoing null hypersurfaces. The resulting
dynamical surface gravity is proportional not only to the frequency decrease of
the outgoing light rays, but also to the acceleration of some special family of
observers. Furthermore, any observer can easily measure it by observing the
apparent redshift of standard light signals falling in from infinity. The fact
that the surface gravity can be expressed as an integral along an ingoing null
line shows that it is always a continuous function along the apparent horizon,
even if there are sudden changes in the matter field density.

We have also seen that the area change of the apparent horizon, which may be
essential in possible thermodynamical interpretations, becomes directly
proportional to the surface gravity only in the stationary limit. On the other
hand, although it is well known that stationary black holes emit Hawking
radiation with temperature proportional to their surface gravity, it is unclear
whether or not a dynamical analogue of this statement can be formulated. There
have been attempts to define dynamical temperature only at the (approximate)
event horizon of the Vaidya spacetime\cite{vate}. If there was discrepancy
between the temperature and the surface gravity, it might be linked with the
non-thermal nature of the Hawking radiation.

Based on the study of the examples in the previous section, we can have a
number of conjectures on the general dynamical behavior of surface gravity in
spherically symmetric spacetimes. It is natural to expect that the surface
gravity of evaporating black holes is always a positive and non decreasing
function of time. This case is especially important, since if similarly to the
quasi-static limit there was a close relation between the temperature of the
(thermal part of the) Hawking radiation and the surface gravity, then this
would be the strongest support in favour of our definitions. Although these
kinds of calculations are extremely difficult to perform in four-dimensional
Einstein theory, it is very encouraging that the generalized surface gravity of
black holes in the exactly solvable two-dimensional dilaton gravity models
(CGHS and RST) is a positive mass independent constant, in accordance with the
Hawking temperature calculations\cite{stro}.

If the matter fields falling into the black hole satisfy the energy conditions,
then according to the classification of Hayward\cite{hayw} (see Appendix), the
apparent horizon is either spacelike outer or timelike inner. From the examples
it seems very likely that spacelike outer horizons always have non-negative
surface gravity which decreases in the outgoing direction in most of the
physically relevant cases. The surface gravity of timelike inner horizons is
probably always a decreasing function of time if the singularity is in the
future. Since the surface gravity is continuous, it is initially positive even
in the inner region. If this inner region is large enough, then the surface
gravity can become negative there.

It is possible that the positivity of $\kappa$ may be proved somehow by the
integral formula (\ref{intka}), if one uses the energy conditions and that
$2M<r$ outside of the horizon. Although equation (\ref{ett}) gives us the
derivative of $\kappa$ in the ingoing null direction, there are no similar
relations for the other directional derivatives. Substituting (\ref{taa}) and
(\ref{evr}) into (\ref{ett}), we get
\begin{equation}
\kappa_{,r}=\frac{8\pi G}{r^2}T_{\theta\theta}-\frac{4\pi F}GT_{rr}+
\frac{2G}r\left(\frac Mr\right)_{,r} \ .
\end{equation}
At the apparent horizon $F=0$ and $r=2M$. Since outside of the horizon $r>2M$,
the third term is negative for spacelike outer and positive for timelike inner
horizons (see Fig\ \ref{figpor}). At the boundary of these two regions, where
the horizon is ingoing null, $\kappa_{,r}$ is exactly the horizon directional
derivative, and its signature is determined only by the signature of the
angular directional pressure $\frac1{r^2}P_{\theta\theta}$. In particular, for
collapsing dust $T_{\theta\theta}=0$ and $\kappa_{,r}=0$. This indicates that
in case of dust collapse the surface gravity takes its maximal value exactly
where the apparent horizon becomes an ingoing null hypersurface.

\acknowledgments

We would like to thank Dr.\ J.\ Soda for suggesting that we should calculate
the surface gravity of black holes in the exactly solvable two-dimensional
dilaton gravity models. One of the authors (G.\ F.) would like to thank the
members of the Cosmology and Gravitation Laboratory of Nagoya University for
their kind hospitality, and acknowledges the support of OTKA grant no.\ T
017176.

\appendix
\section*{}

In this appendix we review some important properties of radial null congruences
in spherically symmetric spacetimes. We mostly follow the approach of
Hayward\cite{hayw}, with the exception that we do not assume that our vector
fields are generated by null foliations. Denote the future directed null vector
fields generating the congruences by $k^\alpha_+$ and $k^\alpha_-$, pointing in
the outgoing and ingoing radial null directions respectively. Define the
normalization function $f$ by  $k^\alpha_+k_{-\alpha}=-e^{-f}$. Because of
spherical symmetry, $k^\alpha_\pm$ are geodesics, although they are not
necessarily affinely parametrized:
\begin{equation}
k^\beta_+k^\alpha_{+;\beta}=b_+k^\alpha_+\ , \qquad
k^\beta_-k^\alpha_{-;\beta}=b_-k^\alpha_-\ .
\end{equation}

Given two intersecting ingoing and outgoing foliations of null hypersurfaces
determined by constant values of $\xi^+$ and $\xi^-$, there are two obvious
ways to define the null vector fields. The first is to set
$k_{\pm\alpha}=-\xi^\mp_{\ ;\alpha}$. Then $b_\pm=0$ and $k^\alpha_\pm$ are
affine geodesics.  The another way is to define the null vector fields, by
\begin{equation}
k^\alpha_\pm=\left(\frac \partial{\partial\xi^\pm}\right)^\alpha\ .
\end{equation}
In this case the inaffinity parameters are
\begin{equation}
b_\pm=-k^\alpha_\pm f_{;\alpha}\ ,
\end{equation}
which can be easily checked in the null coordinate system adapted to the
foliations. Note that in this case $\xi^+_{;\alpha}\xi^{-;\alpha}=-e^f$, in
accordance with Hayward's notation.

The tensor
\begin{equation}
h_{\alpha\beta}=g_{\alpha\beta}+e^f(k_{+\alpha}k_{-\beta}
+k_{-\alpha}k_{+\beta})
\end{equation}
acts as a projection operator into the 2-spheres. For the calculation of the
expansion, shear and twist, we need to evaluate the tensor
$B^\pm_{\alpha\beta}=h^\gamma_\alpha h^\delta_\beta k_{\pm\alpha;\beta}$. Since
$B^\pm_{\alpha\beta}$ is symmetric and its trace-free part vanishes, the twist
and the shear is zero. The expansion is
\begin{equation}
\Theta_\pm=B^{\pm\alpha}_\alpha=\frac2\rho k^\alpha_\pm\rho_{;\alpha}=\frac1
Ak^\alpha_\pm A_{;\alpha}\ ,
\end{equation}
where the area function is $A=4\pi\rho^2$.
Another useful formula relating the Lie derivative of $h_{\alpha\beta}$ to the
expansion is $\Theta_\pm=h^{\alpha\beta}{\cal L}_{k_\pm}h_{\alpha\beta}/2$.

Using the Einstein's equations, one can derive two useful expressions for the
directional derivatives of the expansions. The formula corresponding to the
Raychaudhuri equation is
\begin{equation}
k^\alpha_\pm\Theta_{\pm;\alpha}=-\frac12\Theta^{\ 2}_\pm+b_\pm\Theta_\pm
-8\pi T_{\alpha\beta}k^\alpha_\pm k^\beta_\pm\  ,
\end{equation}
where $T_{\alpha\beta}$ is the stress tensor of the matter fields. The
cross-focusing equation gives the  derivative in the another null direction:
\begin{equation}
k^\alpha_\mp\Theta_{\pm;\alpha}=-\Theta_+\Theta_- -\left(k^\alpha_\mp
f_{;\alpha}+b_\mp\right)\Theta_\pm
-\frac1{\rho^2}e^{-f}+8\pi T_{\alpha\beta}k^\alpha_+ k^\beta_-\  .
\label{crosf}
\end{equation}

If either of the two expansions $\Theta_+$ or $\Theta_-$ vanishes on a sphere
of symmetry, the sphere is called a {\em marginal sphere}. The closure of a
hypersurface foliated by marginal spheres is called a {\em trapping horizon}. A
marginal sphere on the horizon with $\Theta_+=0$ is called {\em future} if
$\Theta_-<0$ and {\em past} if $\Theta_->0$. It is called {\em outer} if
$k^\alpha_-\Theta_{+;\alpha}<0$, and {\em inner} if
$k^\alpha_-\Theta_{+;\alpha}>0$. If the weak energy condition  holds, outer
horizons are spacelike or null, while inner horizons are timelike or null. In
both cases, they are null in the $k^\alpha_+$ direction if and only if
$T_{\alpha\beta}k^\alpha_+k^\beta_+=0$.

\begin{figure}
  \begin{center}
    \leavevmode
    \epsfysize=15pc \epsfbox[121 401 471 605]{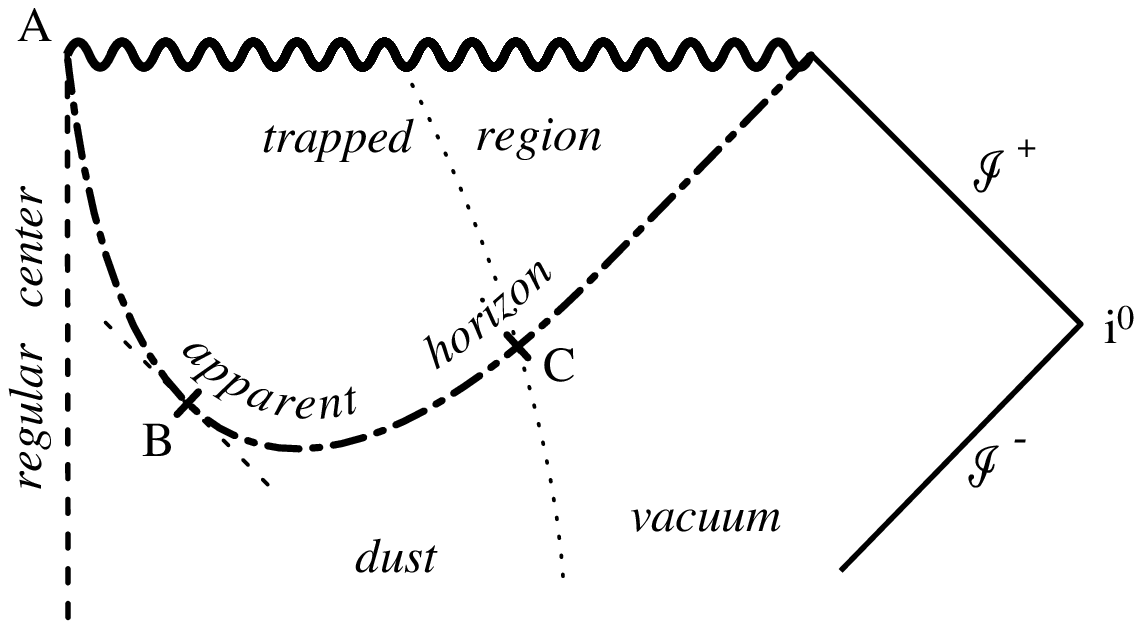}
  \end{center}
\caption{
Conformal diagram describing the collapse of an inhomogeneous dust ball. The
apparent horizon is timelike inner between the points A and B, spacelike outer
between B and C, and null in the vacuum region. If there is a sudden change in
the density, for example at the surface of the ball, then the direction of the
horizon changes non-continuously there. Depending on the the initial density
distribution, a null singularity may appear at the center, and near it the
apparent horizon has to become spacelike outer again.
}\label{figpor}
\end{figure}

The naming `inner' for horizons satisfying $k^\alpha_-\Theta_{+;\alpha}>0$ can
be misleading though. These inner horizons can separate a trapped region from
an asymptotically flat untrapped region. A future horizon which is a smooth
connected hypersurface can be outer in one region and change to be inner in
another region, simply by becoming timelike through ingoing null directions.
For example in certain cases of pressureless dust collapse, the horizon can be
timelike-inner in a region close to the regular center, analogously to the
cosmological horizon in a collapsing universe. Going outwards, this horizon
becomes ingoing-null at a two-sphere, and then it is spacelike-outer.
Asymptotically, in the Scwarzschild region, the horizon becomes null again, but
then in the outgoing direction (see Fig.\ \ref{figpor}).

Following Hayward\cite{hayw}, we define the {\em trapping gravity} of an outer
trapping horizon by the formula
\begin{equation}
\kappa_H=\frac12\sqrt{-e^fk^\alpha_-\Theta_{+;\alpha}}\ .
\end{equation}
Because of the $e^f$ factor, $\kappa_H$ is independent of the product
$k^\alpha_+k_{-\alpha}=-e^{-f}$, giving the same value as if $k^\alpha_-$ was
chosen to  satisfy  $k^\alpha_+k_{-\alpha}=-1$. However if we rescale the null
vectors by a function $a$ as $\tilde k^\alpha_+=ak^\alpha_+$, $\tilde
k^\alpha_-=k^\alpha_-/a$, keeping $e^f$ unchanged, $\Theta_+$ becomes
$\tilde\Theta_+=a\Theta_+$, and $\kappa_H$ turns into
\begin{equation}
\tilde\kappa_H=\frac12\sqrt{-e^fk^\alpha_-\Theta_{+;\alpha}-\frac1a
e^f\Theta_+k^\alpha_-a_{;\alpha}}\ .
\end{equation}
This shows that $\kappa_H$ is invariantly defined only on the trapping horizon,
where $\Theta_+=0$.

Using the cross-focusing equation (\ref{crosf}), the trapping gravity of the
horizon can be expressed in a local way by the radius function $\rho$ and  the
stress tensor of the matter fields:
\begin{equation}
\kappa_H=\frac12\sqrt{\frac1{\rho^2}-8\pi e^f T_{\alpha\beta}k^\alpha_+
k^\beta_-}\ .
\end{equation}
In the vacuum case we get the familiar $\frac1{2\rho}$ value, agreeing with
surface gravity of the Schwarzschild solution. While $\kappa_H$ is defined on
the trapping horizon of dynamical spacetimes, the surface gravity is defined on
the event horizon of stationary solutions.  For stationary spacetimes the two
kind of horizons coincide. However, in general, the value of the trapping
gravity $\kappa_H$ is different from the value of the surface gravity. This is
the case even for the Reissner-Nordstr\"om solution, the surface gravity of
which is
\begin{equation}
\kappa=\frac1{r^2_h}\sqrt{m^2-e^2}
=\frac1{2r_h}\left(1-\frac{e^2}{r^2_h}\right)\ ,
\end{equation}
where $m$ is the mass, $e$ is the charge parameter, and $r_h=m+\sqrt{m^2-e^2}$.
The value of the trapping gravity is
\begin{equation}
\kappa_H=\frac1{2r^2_h}\sqrt{r_h^2-e^2}\ .
\end{equation}
For fixed $m$, $\kappa$ is a monotonically decreasing function of $e$, while
$\kappa_H$ is not monotonic. $\kappa=\kappa_H$ only for $e=0$ and $e=m$.

Working in the null coordinate system (\ref{nullco}) where $r=\rho$, we can
choose
\begin{equation}
k_+^\alpha=\left(1,\frac F{2G},0,0\right)\ ,\qquad
k_-^\alpha=\left(0,-1,0,0\right)\ .
\end{equation}
Then $e^{-f}=G$ and $\Theta_+=\frac F{rG}$. Since $F=0$ on the horizon,
\begin{equation}
\kappa_H=\frac1{2G}\sqrt{\frac1{r_h} F_{,r}} \ .
\end{equation}
Comparing with (\ref{frkg}) and (\ref{kadt}), we obtain the relation between
our generalized surface gravity $\kappa$, the trapping gravity $\kappa_H$, and
the temperature term $\Theta$ in equation (\ref{fila}):
\begin{equation}
\kappa+4\pi\int^\infty_{r_h}\limits r\frac{\partial T_{rr}}{\partial
v}dr=2r_hG\kappa_H^2=8\pi\Theta\ , \label{conn}
\end{equation}
where the integral is calculated along a constant $v$ ingoing null line. We can
see that the surface gravity agrees with the trapping gravity only in some
special cases. This happens for example for the Vaidya spacetime, where $G=1$
and $T_{rr}=0$.

\vspace{22mm}

\begin{references}

\bibitem{viss} M. Visser, Phys. Rev. D {\bf46}, 2445 (1992)

\bibitem{hisc} W. A. Hiscock, Phys. Rev. D {\bf40},1336 (1989)

\bibitem{haji} P.Hajicek, Phys. Rev. D {\bf36},1065 (1987)

\bibitem{coll} W. Collins, Phys. Rev. D {\bf45}, 495 (1991)

\bibitem{hayw} S. A. Hayward, Phys. Rev. D {\bf49}, 6467 (1994)

\bibitem{cart} B. Carter, in {\it General relativity: An Einstein Centenary
Survey,} edited by S. W. Hawking and W. Israel (Cambridge University Press,
Cambridge, England, 1979)

\bibitem{haha} S. W. Hawking and J. B. Hartle, Commun. Math. Phys. {\bf27}, 283
(1972)

\bibitem{koda} H. Kodama, Prog. Theor. Phys. {\bf63} 1217 (1980)

\bibitem{vaid} B. T. Sullivan and W. Israel, Phys. Lett. {\bf79A}, 371 (1980)

\bibitem{robe} M. D. Roberts, Gen. Rel. Grav. {\bf21}, 907 (1989);  Y. Oshiro,
K. Nakamura and A. Tomimatsu, Prog. Theor. Phys. {\bf91} 1265 (1994)

\bibitem{olch} H. P. Oliveira and E. S. Cheb-Terrab, gr-qc/9512010

\bibitem{cghs} C. G. Callan, S. B. Giddings, J. A. Harvey and A. Strominger,
Phys. Rev. D {\bf45}, R1005 (1992)

\bibitem{rust} J. G. Russo, L. Susskind and L. Thorlacius, Phys. Rev. D
{\bf47}, 533 (1993)

\bibitem{vate} R. Balbinot, Class. Quantum Grav. {\bf1}, 573 (1984);  S. Kim,
E. Choi, S. K. Kim and J. Yang, Phys. Lett. A {\bf141}, 238 (1989)

\bibitem{stro} A. Strominger, hep-th/9501071

\end{references}
\end{document}